\documentclass[12pt]{article}%
\usepackage{amsmath}
\usepackage{amsfonts}
\usepackage{amssymb}
\usepackage{graphicx}%
\providecommand{\U}[1]{\protect\rule{.1in}{.1in}}
\newcommand{\bfr}{\begin{flushright}}
\newcommand{\efr}{\end{flushright}}
\newcommand{\bc}{\begin{center}}
\newcommand{\ec}{\end{center}}
\newcommand{\ben}{\begin{enumerate}}
\newcommand{\een}{\end{enumerate}}

\newcommand{\be}{\begin{equation}}
\newcommand{\ee}{\end{equation}}
\newcommand{\ba}{\begin{array}}
\newcommand{\ea}{\end{array}}
\def\6{\partial}
\begin{document}

\title{\textbf{Noncommutative fluid dynamics} \\\textbf{in the K\"{a}hler parametrization} }
\author{L. Holender$^{a,b}$\thanks{email: holender@ufrrj.br},\, M. A. Santos$^{b}%
$\thanks{email: masantos@cce.ufes.br},\, M. T. D. Orlando$^{b}$\thanks{email:
mtdorlando@terra.com.br}\, and I. V. Vancea$^{a}$\thanks{email:
ionvancea@ufrrj.br}\\$^{a}$\emph{{\small Grupo de F{\'{\i}}sica Te\'{o}rica e Matem\'{a}tica
F\'{\i}sica, Departamento de F\'{\i}sica,}} \\\emph{{\small Universidade Federal Rural do Rio de Janeiro (UFRRJ),}} \\\emph{{\small Cx. Postal 23851, BR 465 Km 7, 23890-000 Serop\'{e}dica - RJ,
Brasil }} \\$^{b}$\emph{{\small Departamento de F\'{\i}sica e Qu\'{\i}mica,}} \\\emph{{\small Universidade Federal do Esp\'{\i}rito Santo (UFES),}} \\\emph{{\small Avenida Fernando Ferarri S/N - Goiabeiras, 29060-900 Vit\'{o}ria
- ES, Brasil}} }
\date{14 September 2011}
\maketitle

\thispagestyle{empty}
%\pagestyle{empty}

%\begin{center}
%\rule{15cm}{0.01cm}
%\end{center}

\abstract{In this paper, we propose a first order action functional for a large class of systems that generalize
the relativistic perfect fluids in the K\"{a}hler parametrization to noncommutative spacetimes.
The noncommutative action is parametrized by two arbitrary functions $K(z,\bar{z})$ and $f(\sqrt-{j^2})$ that
depend on the fluid potentials and represent the generalization of the K\"{a}hler potential of the complex surface
parametrized by $z$ and $\bar{z}$, respectively, and the characteristic function of each model.
We calculate the equations of motion for the fluid potentials and the energy-momentum tensor in the first order
in the noncommutative parameter. The density current does not receive any noncommutative corrections and
it is conserved under the action of the commutative generators $P_{\mu}$ but the energy-momentum
tensor is not. Therefore, we determine the set of constraints under which the energy-momentum tensor is
divergenceless. Another set of constraints on the fluid potentials is obtained from the requirement of the
invariance of the action under the generalization of the volume preserving transformations of the noncommutative
spacetime. We show that the proposed action describes noncommutative fluid models by casting the energy-momentum tensor
in the familiar fluid form and identifying the corresponding energy and momentum densities. In the commutative limit,
they are identical to the corresponding quantities of the relativistic perfect fluids. The energy-momentum tensor
contains a dissipative term that is due to the noncommutative spacetime and vanishes in the commutative limit.
Finally, we particularize the theory to the case when the complex fluid potentials are characterized by a function $K(z,\bar{z})$
that is a deformation of the complex plane and show that this model has important common features with the commutative fluid such as
infinitely many conserved currents and a conserved axial current that in the commutative case is associated
to the topologically conserved linking number.}

%\begin{center}
%\rule{15cm}{0.01cm}
%\end{center}
\vfill

%\begin{flushright}
%{\footnotesize \textbf{DEFIS-ICE-UFRRJ/2011 \hspace{0.5cm} TH-PHYS/02} }
%\end{flushright}

%-------------------------------------------------------------------------------
\newpage\pagestyle{plain} \pagenumbering{arabic}

\section{Introduction}

The formulation of a theory of the noncommutative fluids was motivated
initially by the observation that the abelian noncommutative Chern-Simons
theory at level $n$ is equivalent to the Laughlin theory at level $1/n$
\cite{Susskind:2001fb,Bahcall:1991an} thus establishing a connection among the
theories of noncommutative fields, fluid dynamics, quantum Hall effect and the
matrix theory. The connection between the fractional quantum Hall effect and
the noncommutative field theory has been subsequently studied for the Haldane
model in \cite{ElRhalami:2001xf} while the noncommutative fluid model from
\cite{Susskind:2001fb} was used to determine the density fluctuations in
\cite{Barbon:2001dw} and the topological order of the fractional Hall effect
in \cite{Barbon:2007zz}(see for a review \cite{Polychronakos:2007df}). A
different motivation for the study of the noncommutative fluids is given by
the fact that the volume preserving transformations leave invariant the
structure of noncommutative configuration spaces as well as the equations of
motion of the nonabelian Lagrangian fluids
\cite{Jackiw:2002pn,Jackiw:2002tw,Jackiw:2003dw,Jackiw:2004nm}. More recently,
different fluid models have appeared in the context of $U(1)$ gauge fields in
curved noncommutative spaces \cite{Alavi:2006sr} and in the study of the
cosmological perturbations of the perfect fluid \cite{DeFelice:2009bx}. In
\cite{Marcial:2010zza}, it was proposed a generalized symplectic structure of
two models of irrotational and rotational noncommutative nonrelativistic
fluids, respectively.

When studying the noncommutative fluids, it is certainly important to
investigate models that reduce to relativistic fluids in the limit of
commutative spacetime. This task is facilitated by the existence of a
formulation in terms of the action functional of a large class of relativistic
(perfect) fluids. In this formulation the fluid degrees of freedom that enter
a first-order Lagrangian are given by the fluid potentials in either the
(real) Clebsch parametrization \cite{lect-jackiw} or the (complex) K\"{a}hler
parametrization \cite{Nyawelo:2003bv}. Although a proof of the equivalence of
the two parametrizations is missing, it is known that both of them remove the
obstruction to define a consistent Lagrangian which is due to the Chern-Simons
term that is necessary in order to describe the nonzero vorticity and can be
generalized to include the supersymmetry \cite{Nyawelo:2003bv,Jackiw:2000cc}.
The complex parametrization of the fluid potentials has two interesting
properties. Firstly, there are infinitely many conserved charges for the
non-singular K\"{a}hler potentials that characterize a geodesically complete
complex manifold. Secondly, the Hamiltonian dynamics is governed by a set of
simple second-class constraints among the fluid degrees of freedom. In
particular, that Hamiltonian structure of the constraints has permitted a
detailed analysis of the metafluid dynamics in \cite{Baleanu:2004sc}, the
formulation of the conformal fluids in \cite{Jarvis:2005hp} and the
quantization of a large class of non-supersymmetric fluids in
\cite{Holender:2008qj}. Also, the K\"{a}hler parametrization has been used to
formulate the supersymmetric hydrodynamics in \cite{Nyawelo:2003bw} and to
construct the Navier-Stokes equations from the AdS/CFT and fluid
correspondence in \cite{Grassi:2011wt}.

In this paper, we propose an action for the noncommutative fluid that
generalizes the action of the relativistic fluid in the K\"{a}hler
parametrization to the noncommutative space $M_{\lambda}$ defined by the
relations
\begin{equation}
\left[  x^{\mu},x^{\nu}\right]  =i\lambda^{\mu\nu}, \label{nc-rel}%
\end{equation}
where $\mu,\nu=0,1,2,3$ and $\lambda^{\mu\nu}$ is a constant antisymmetric
matrix. Our action reduces to the previous action from \cite{Nyawelo:2003bv}
in the commutative limit $\lambda^{\mu\nu}\rightarrow0$. The noncommutative
action is not Poincar\'{e} invariant since the relevant group in the general
noncommutative space given by the relation (\ref{nc-rel}) is the volume
preserving group rather than a deformation of the Poincar\'{e} group. We
determine a set of constraints on the fluid potentials such that the
Lagrangian is invariant under the volume preserving group. By choosing the
commuting conjugate operators $P_{\mu}$ to $x^{\mu}$ we show that, contrary to
the commutative case, the energy-momentum tensor of the noncommutative fluid
is not divergenceless under the action of $P_{\mu}$'s. However, we are able to
determine a set of constraints for the fluid potentials under which the
energy-momentum tensor is conserved under the commutative translations.

This paper is organized as follows. In the next section we propose the action
of a large class of noncommutative fluids parametrized by the generalizations
of the K\"{a}hler potential and an arbitrary function on the fluid currents
that characterizes particular models from this class. Also, we derive the
equations of motion at first order in $\lambda_{\mu\nu}$. In section 3, we
derive the energy-momentum tensor and the equation of state. In the
commutative limit, they take the form of the corresponding equations of the
relativistic perfect fluid. In section 4 we determine the constraints on the
fluid potentials under which the noncommutative Lagrangian is invariant at
zeroth and first order in the noncommutative parameter. \ In section 5 we
present a simpler model which generalizes the fluid potentials on the complex
plane. We show that in this model there are infintely many conserved currents
as in the commutative case, which makes the model particulary interesting
because, in general, the generalizations of the fluid do not have this
property. The last section is devoted to discussions.

\section{Noncommutative fluid action}

The class of relativistic perfect fluids on the four-dimensional Minkowski
space $M$ can be described in terms of the scalar potentials $\{\theta
(x),z(x),\bar{z}(x)\}$ which are smooth functions from $\mathcal{C}^{\infty
}(M)=\{f:M\rightarrow%
%TCIMACRO{\U{2102} }%
%BeginExpansion
\mathbb{C}
%EndExpansion
\}$. The potential $\theta\left(  x\right)  $ is purely real while the fields
$z(x)$ and $\bar{z}(x)$ are complex conjugate to each other, respectively. The
class is parametrized by two arbitrary functions: $K(z,\bar{z})$ that is the
K\"{a}hler potential associated to the two dimensional manifold of coordinates
$z$ and $\bar{z}$ and $f(\rho)$ which depends on the local fluid density
$\rho$. The relativistic fluid is characterized by the equations of state that
involve the local pressure $p$ and the energy density $\varepsilon$,
respectively. The dynamics conserves the energy-momentum tensor $T_{\mu\nu}$
and the fluid density current $j^{\mu}$\ and can be derived from a first order
Lagrangian functional in the potentials\footnote{The metric on the Minkowski
space has the signature $\left(  -,+,+,+\right)  $. The current four-vector is
defined as $j^{\mu}=\rho u^{\mu}$ where $u^{\mu}=dx^{\mu}/d\tau$ is the
velocity four-vector and $u^{\mu}u_{\mu}=-1$.}. The Lagrangian has two more
symmetries: the parametrization of the fluid potentials which lead to the
conservation of infinitely many two dimensional currents $J_{\mu}$ and the
axial symmetry which leads to the conservation of the topological charge
$\omega$ that describes the linking number of the vortices formed in the fluid
\cite{Nyawelo:2003bv,Holender:2008qj}.

Consider the noncommutative space $M_{\lambda}$ with the algebra of complex
function $\mathcal{F}(M_{\lambda})$. A well know property \cite{Connes:1994yd}
is that this structure is isomorphic to the algebra $\left(  \mathcal{C}%
^{\infty}(M),\ast\right)  $ where $\ast:\mathcal{C}^{\infty}(M)\times
\mathcal{C}^{\infty}(M)\rightarrow\mathcal{C}^{\infty}(M)$ is the Moyal
product defined as
\begin{equation}
f\ast g=fe^{\frac{i}{2}\lambda_{\mu\nu}\overleftarrow{\partial}_{\mu
}\overrightarrow{\partial}_{\nu}}g. \label{Moyal-prod}%
\end{equation}
We take for the tangent space mapping%
\begin{equation}
\left[  \partial_{\mu},\partial_{\nu}\right]  =0. \label{tang-space}%
\end{equation}
Since the algebra of functions contains the same objects with the usual dot
product replaced by star product, the perfect fluid is still characterized by
its potentials $\{\theta(x),z(x),\bar{z}(x)\}$ with the interaction given by
the star multiplication which could possibly affect the physical properties of
the system. The action functional of the noncommutative fluid that generalizes
the commutative action from \cite{Nyawelo:2003bv}\ is given by the following
relation%
\begin{equation}
S\left[  j^{\mu},\theta,z,\bar{z}\right]  =\int d^{4}x\left[  -j^{\mu}%
\ast\left(  \partial_{\mu}\theta+i\partial_{z}K\ast\partial_{\mu}%
z-i\partial_{\bar{z}}K\ast\partial_{\mu}\bar{z}\right)  \right]  -f\left(
\sqrt{-j^{\mu}\ast j_{\mu}}\right)  . \label{nonc-action}%
\end{equation}
The Lagrangian from (\ref{nonc-action})\ describes a large class of 
noncommutative fluids
parametrized by the arbitrary functions $K(z,\bar{z})$ and $f\left(
\sqrt{-j^{\mu}\ast j_{\mu}}\right)  $. In what follows, we are going to study
the action (\ref{nonc-action}) for a general noncommutative field $j^{\mu}$
until section 4 where we will investigate the consequences of the
generalization of the relation $\rho u^{\mu}$ to the noncommutative theory. In
general, $K(z,\bar{z})$ is not associated to a noncommutative K\"{a}hler
manifold which can be viewed as a deformation quantization of a K\"{a}hler
manifold (see e. g. \cite{Omori:1997,Reshetikhin:2000,Schlichenmaier:2010ui}.
However, the commutative sector of $K(z,\bar{z})$ is the K\"{a}hler potential
on the commutative sector of the $(z,\bar{z})$ manifold. In what follows, we
make the simplifying truncation of the partial derivatives of the generalized
K\"{a}hler potential at zero order in $\lambda_{\mu\nu}$ which allows one to
apply the Leibnitz rule. If higher orders in the noncommutative parameter are
considered, the Leibniz rule does not generally hold. The function $f\left(
\sqrt{-j^{\mu}\ast j_{\mu}}\right)  $ should coincide with $f\left(
\sqrt{-j^{\mu}j_{\mu}}\right)  $ in the commutative limit $\lambda_{\mu\nu
}\rightarrow0$. In this way, it is established the correspondence principle
between the noncommutative perfect fluids given by the action
(\ref{nonc-action}) and the commutative perfect fluids studied in
\cite{Nyawelo:2003bv,Holender:2008qj}. For small values of $\lambda_{\mu\nu}$
the linearized Lagrangian from the equation (\ref{nonc-action}) takes the form%
\begin{align}
\mathcal{L}\left[  j^{\mu},\theta,z,\bar{z}\right]  =  &  -j^{\mu}\left(
\partial_{\mu}\theta+i\partial_{z}K\cdot\partial_{\mu}z-i\partial
_{\overline{z}}K\cdot\partial_{\mu}\overline{z}\right) \nonumber\\
&  +\frac{1}{2}\lambda^{\alpha\beta}j^{\mu}\left(  \partial_{\alpha}%
\partial_{z}K\cdot\partial_{\beta}\partial_{\mu}z-\partial_{\alpha}%
\partial_{\overline{\bar{z}}}K\cdot\partial_{\beta}\partial_{\mu}\overline
{z}\right) \nonumber\\
&  -\frac{i}{2}\lambda^{\alpha\beta}\partial_{\alpha}j^{\mu}\cdot
\partial_{\beta}\left(  \partial_{\mu}\theta+i\partial_{z}K\cdot\partial_{\mu
}z-i\partial_{\overline{z}}K\cdot\partial_{\mu}\overline{z}\right) \nonumber\\
&  -f\left(  \sqrt{-j^{2}-\frac{i}{2}\lambda^{\alpha\beta}\partial_{\alpha
}j^{\mu}\partial_{\beta}j_{\mu}}\right)  . \label{nonc-lagrangian}%
\end{align}
A first difference to be noted between the commutative and the noncommutative
fluids is that the current $j^{\mu}$ is propagating in the noncommutative
case. Also, even in the lowest order in the noncommutative parameter, the
Lagrangian contains higher order derivatives in the fields. The Euler-Lagrange
equations of motion can be obtained in the usual way by imposing the
invariance of the action (\ref{nonc-action}) under infinitesimal variations of
the fields with vanishing boundary conditions for the fields and the
derivatives. As can be seen from (\ref{nonc-action}), the equations of motion
have the general form%
\begin{equation}
\frac{\delta\mathcal{L}}{\delta\phi}=\frac{\partial\mathcal{L}}{\partial\phi
}-\frac{\partial}{\partial x^{\alpha}}\left(  \frac{\partial\mathcal{L}%
}{\partial\left(  \partial_{\alpha}\phi\right)  }\right)  +\frac{\partial^{2}%
}{\partial x^{\alpha}\partial x^{\beta}}\left(  \frac{\partial\mathcal{L}%
}{\partial\left(  \partial_{\alpha\beta}^{2}\phi\right)  }\right)  =0.
\label{gen-eq-motion}%
\end{equation}
By calculating (\ref{gen-eq-motion}) for the scalar potential $\theta(x)$, one
can easily show that%
\begin{equation}
\partial_{\mu}j^{\mu}=0. \label{eq-mot-theta}%
\end{equation}
The equation of motion of the current $j^{\mu}$ takes the following form%
\begin{align}
f^{\prime}\frac{j_{\mu}}{\sqrt{-j^{2}-\frac{i}{2}\lambda^{\alpha\beta}%
\partial_{\alpha}j^{\nu}\partial_{\beta}j_{\nu}}}  &  =\left(  \partial_{\mu
}\theta+i\partial_{z}K\cdot\partial_{\mu}z-i\partial_{\overline{z}}%
K\cdot\partial_{\mu}\overline{z}\right) \nonumber\\
&  -\frac{1}{2}\lambda^{\alpha\beta}\left(  \partial_{\alpha}\partial
_{z}K\cdot\partial_{\beta}\partial_{\mu}z-\partial_{\alpha}\partial_{\bar{z}%
}K\cdot\partial_{\beta}\partial_{\mu}\overline{z}\right)  . \label{eq-mot-j}%
\end{align}
Here, $f^{\prime}$ denotes the derivative of $f$ with respect to its variable.
The equation of motion of the potential $z(x)$ can be obtained in the same
way from the equation (\ref{gen-eq-motion}). After some algebra, one can show
that it has the following form%
\begin{align}
&  -ij^{\mu}\left(  \partial_{zz}^{2}K\cdot\partial_{\mu}z-\partial_{z\bar{z}%
}^{2}K\cdot\partial_{\mu}\bar{z}\right)  +i\partial_{\mu}\left(  j^{\mu
}\partial_{z}K\right)  +\frac{1}{2}\lambda^{\alpha\beta}j^{\mu}\left(
\partial_{z}\partial_{\alpha}\partial_{z}K\cdot\partial_{\beta}\partial_{\mu
}z-\partial_{z}\partial_{\alpha}\partial_{\overline{\bar{z}}}K\cdot
\partial_{\beta}\partial_{\mu}\overline{z}\right) \nonumber\\
&  +\frac{1}{2}\lambda^{\alpha\beta}\partial_{\alpha}j^{\mu}\cdot\left(
\partial_{z}\partial_{\beta}\partial_{z}K\cdot\partial_{\mu}z-\partial
_{z}\partial_{\beta}\partial_{\overline{z}}K\cdot\partial_{\mu}\overline
{z}\right)  +\frac{1}{2}\lambda^{\alpha\beta}\partial_{\alpha}\left[  j^{\mu
}\left(  \partial_{zz}^{2}K\cdot\partial_{\beta}\partial_{\mu}z-\partial
_{z\bar{z}}^{2}K\cdot\partial_{\beta}\partial_{\mu}\bar{z}\right)  \right]
\nonumber\\
&  +\frac{1}{2}\lambda^{\alpha\beta}\partial_{\beta}\left[  \partial_{\alpha
}j^{\mu}\left(  \partial_{zz}^{2}K\cdot\partial_{\mu}z-\partial_{z\bar{z}}%
^{2}K\cdot\partial_{\mu}\bar{z}\right)  \right]  +\frac{1}{2}\lambda
^{\alpha\beta}\partial_{\beta\mu}^{2}\left(  j^{\mu}\partial_{\alpha}%
\partial_{z}K\right) \nonumber\\
&  =0. \label{eq-mot-z}%
\end{align}
The equation of motion of $\bar{z}(x)$ can be obtained from (\ref{eq-mot-z})
by replacing the appropriate derivative with respect to $z$ by derivative with
respect to $\bar{z}$ or by using (\ref{gen-eq-motion}). By either way, the
result is%
\begin{align}
&  -ij^{\mu}\left(  \partial_{\bar{z}z}^{2}K\cdot\partial_{\mu}z-\partial
_{\bar{z}\bar{z}}^{2}K\cdot\partial_{\mu}\bar{z}\right)  +i\partial_{\mu
}\left(  j^{\mu}\partial_{\bar{z}}K\right)  +\frac{1}{2}\lambda^{\alpha\beta
}j^{\mu}\left(  \partial_{\bar{z}}\partial_{\alpha}\partial_{z}K\cdot
\partial_{\beta}\partial_{\mu}z-\partial_{\bar{z}}\partial_{\alpha}%
\partial_{\overline{\bar{z}}}K\cdot\partial_{\beta}\partial_{\mu}\overline
{z}\right) \nonumber\\
&  +\frac{1}{2}\lambda^{\alpha\beta}\partial_{\alpha}j^{\mu}\cdot\left(
\partial_{\bar{z}}\partial_{\beta}\partial_{z}K\cdot\partial_{\mu}%
z-\partial_{\bar{z}}\partial_{\beta}\partial_{\overline{z}}K\cdot\partial
_{\mu}\overline{z}\right)  +\frac{1}{2}\lambda^{\alpha\beta}\partial_{\alpha
}\left[  j^{\mu}\left(  \partial_{z\bar{z}}^{2}K\cdot\partial_{\beta}%
\partial_{\mu}z-\partial_{\bar{z}\bar{z}}^{2}K\cdot\partial_{\beta}%
\partial_{\mu}\bar{z}\right)  \right] \nonumber\\
&  +\frac{1}{2}\lambda^{\alpha\beta}\partial_{\beta}\left[  \partial_{\alpha
}j^{\mu}\left(  \partial_{z\bar{z}}^{2}K\cdot\partial_{\mu}z-\partial_{\bar
{z}\bar{z}}^{2}K\cdot\partial_{\mu}\bar{z}\right)  \right]  +\frac{1}%
{2}\lambda^{\alpha\beta}\partial_{\beta\mu}^{2}\left(  j^{\mu}\cdot
\partial_{\alpha}\partial_{\bar{z}}K\right) \nonumber\\
&  =0. \label{eq-mot-z-bar}%
\end{align}
Note that the derivatives with respect to the spacetime coordinates do not
commute with the derivatives with respect to the complex fields $z$ and
$\bar{z}$, respectively. The first of the equations of motion
(\ref{eq-mot-theta}) has a simple interpretation. It shows that the current
$j^{\mu}$ is invariant under the transformations generated by the operators
$P_{\mu}=\partial_{\mu}$. This equation does not receive any noncommutative
corrections and it is in agreement with the generalization of the translation
group defined by the equation (\ref{tang-space}). The remaining equations of
motion do not have such simple interpretation but more algebra shows that they
reduce to the corresponding equations in the commutative limit. In particular,
the equations (\ref{eq-mot-z}) and (\ref{eq-mot-z-bar}) do not imply any
longer that there are infinitely many conserved currents associated to the
reparametrization invariance of any K\"{a}hler surface. We will return to this
point in section 5.

\section{Energy-momentum tensor}

The class of perfect relativistic fluids in the Minkowski spacetime which are
generalized to the noncommutative spacetime by the action (\ref{nonc-action}%
)\ are characterized by the divergenceless density current and the
divergenceless energy-momentum tensor. These properties are related to the
equations of motion of the fluid and with the translation invariance of the
Lagrangian. As we have seen in the previous section, the density current of
the noncommutative fluid is divergenceless and, by identifying the generators
of the translations with the derivatives $\partial_{\mu}$, it is related to
the translation invariance, too.

The energy-momentum tensor of the noncommutative fluid can be defined by
coupling it with a $c$ - number metric tensor $g_{\mu\nu}(x)$ and by taking
the functional derivative of the action with respect to the metric. In this
way we obtain the relation%
\begin{align}
T_{\mu\nu}  &  =\eta_{\mu\nu}\left[  -j^{\gamma}\left(  \partial_{\gamma
}\theta+i\partial_{z}K\cdot\partial_{\gamma}z-i\partial_{\overline{z}}%
K\cdot\partial_{\gamma}\overline{z}\right)  +\frac{1}{2}\lambda^{\alpha\beta
}j^{\gamma}\left(  \partial_{\alpha}\partial_{z}K\cdot\partial_{\beta}%
\partial_{\gamma}z-\partial_{\alpha}\partial_{\bar{z}}K\cdot\partial_{\beta
}\partial_{\gamma}\bar{z}\right)  \right. \nonumber\\
&  \left.  -\frac{i}{2}\lambda^{\alpha\beta}\partial_{\alpha}j^{\gamma}%
\cdot\partial_{\beta}\left(  \partial_{\gamma}\theta+i\partial_{z}%
K\cdot\partial_{\gamma}z-i\partial_{\overline{z}}K\cdot\partial_{\gamma
}\overline{z}\right)  -f\left(  \sqrt{-j^{\mu}j_{\mu}+\frac{i}{2}%
\lambda^{\alpha\beta}\partial_{\alpha}j^{\mu}\cdot\partial_{\beta}j_{\mu}%
}\right)  \right] \nonumber\\
&  +2j_{\mu}\left(  \partial_{\nu}\theta+i\partial_{z}K\cdot\partial_{\nu
}z-i\partial_{\overline{z}}K\cdot\partial_{\nu}\overline{z}\right)
-\lambda^{\alpha\beta}j_{\mu}\left(  \partial_{\alpha}\partial_{z}%
K\cdot\partial_{\beta}\partial_{\nu}z-\partial_{\alpha}\partial_{\bar{z}%
}K\cdot\partial_{\beta}\partial_{\nu}\bar{z}\right) \nonumber\\
&  +i\lambda^{\alpha\beta}\partial_{\alpha}j_{\mu}\cdot\partial_{\beta}\left(
\partial_{\nu}\theta+i\partial_{z}K\cdot\partial_{\nu}z-i\partial
_{\overline{z}}K\cdot\partial_{\nu}\overline{z}\right)  -f^{\prime}%
\cdot\left(  j_{\mu}j_{\nu}+\frac{i}{2}\lambda^{\alpha\beta}\partial_{\alpha
}j_{\mu}\cdot\partial_{\beta}j_{\nu}\right)  . \label{em-tensor}%
\end{align}
In general, the divergence of the energy-momentum tensor (\ref{em-tensor})
will not vanish. In order for this to happen, one has to impose constraints on
the fields. It can be shown that by using the equations of motion
(\ref{const-lambda1-h1}) and (\ref{const-lambda1-h2}) the energy-momentum
tensor is divergenceless for the solutions of the following constraints
\begin{align}
&  \partial_{\nu}\left(  f^{\prime}\frac{j^{\mu}j_{\mu}}{\sqrt{-j^{2}-\frac
{i}{2}\lambda^{\beta\gamma}\partial_{\beta}j^{\nu}\partial_{\gamma}j_{\nu}}%
}-f\right)  -j_{\mu}\partial^{\mu}\left(  f^{\prime}\frac{j_{\nu}}%
{\sqrt{-j^{2}-\frac{i}{2}\lambda^{\beta\gamma}\partial_{\beta}j^{\nu}%
\partial_{\gamma}j_{\nu}}}\right)  =0,\label{cons-em-tensor-1}\\
&  \partial_{\nu}\partial_{\alpha}j^{\mu}\cdot\partial_{\beta}\left(
\partial_{\mu}\theta+i\partial_{z}K\cdot\partial_{\mu}z-i\partial
_{\overline{z}}K\cdot\partial_{\mu}\overline{z}\right) \nonumber\\
&  -\partial_{\alpha}j_{\mu}\cdot\partial^{\mu}\partial_{\beta}\left(
\partial_{\nu}\theta+i\partial_{z}K\cdot\partial_{\nu}z-i\partial
_{\overline{z}}K\cdot\partial_{\nu}\overline{z}\right)  +\partial_{\alpha
}j_{\mu}\cdot\partial^{\mu}\partial_{\beta}j_{\nu}=0. \label{cons-em-tensor-2}%
\end{align}
In the form given by the equation (\ref{em-tensor}), it is unclear how the
commutative perfect fluid is generalized to the noncommutative space. In order
to make the relationship between the two more transparent, we take for the
noncommutative $j^{\mu}$ the following natural generalization of the current%
\begin{equation}
j^{\mu}=\rho\ast u^{\mu}, \label{nonc-j}%
\end{equation}
where $u^{\mu}=dx^{\mu}/d\tau$ does depend on $\tau$ only. Then it is easy to
verify that%
\begin{equation}
f\left(  \sqrt{-j^{\mu}j_{\mu}+\frac{i}{2}\lambda^{\alpha\beta}\partial
_{\alpha}j^{\mu}\cdot\partial_{\beta}j_{\mu}}\right)  =f\left(  \sqrt{-j^{\mu
}j_{\mu}}\right)  . \label{nonc-f}%
\end{equation}
By performing the corresponding simplification and by using the equation of
motion of \ $j^{\mu}$ (\ref{const-lambda1-h1}), one can show that the
energy-momentum tensor has the following form%
\begin{equation}
T_{\mu\nu}=\eta_{\mu\nu}p(\lambda)+\left[  \varepsilon(\lambda)+p(\lambda
)\right]  u_{\mu}u_{\nu}+i\lambda^{\alpha\beta}\partial_{\alpha}\rho\cdot
u_{\mu}\partial_{\beta}\left(  \partial_{\nu}\theta+i\partial_{z}%
K\cdot\partial_{\nu}z-i\partial_{\overline{z}}K\cdot\partial_{\nu}\overline
{z}\right)  , \label{em-tensor-fluid}%
\end{equation}
where%
\begin{align}
p(\lambda)=  &  \rho f^{\prime}-f-j^{\gamma}\left(  \partial_{\gamma}%
\theta+i\partial_{z}K\cdot\partial_{\gamma}z-i\partial_{\overline{z}}%
K\cdot\partial_{\gamma}\overline{z}\right)  +\frac{1}{2}\lambda^{\alpha\beta
}j^{\gamma}\left(  \partial_{\alpha}\partial_{z}K\cdot\partial_{\beta}%
\partial_{\gamma}z-\partial_{\alpha}\partial_{\bar{z}}K\cdot\partial_{\beta
}\partial_{\gamma}\bar{z}\right) \nonumber\\
&  -\frac{i}{2}\lambda^{\alpha\beta}\partial_{\alpha}j^{\gamma}\cdot
\partial_{\beta}\left(  \partial_{\gamma}\theta+i\partial_{z}K\cdot
\partial_{\gamma}z-i\partial_{\overline{z}}K\cdot\partial_{\gamma}\overline
{z}\right)  ,\label{nonc-p}\\
\varepsilon(\lambda)=  &  f+j^{\gamma}\left(  \partial_{\gamma}\theta
+i\partial_{z}K\cdot\partial_{\gamma}z-i\partial_{\overline{z}}K\cdot
\partial_{\gamma}\overline{z}\right)  -\frac{1}{2}\lambda^{\alpha\beta
}j^{\gamma}\left(  \partial_{\alpha}\partial_{z}K\cdot\partial_{\beta}%
\partial_{\gamma}z-\partial_{\alpha}\partial_{\bar{z}}K\cdot\partial_{\beta
}\partial_{\gamma}\bar{z}\right) \nonumber\\
&  +\frac{i}{2}\lambda^{\alpha\beta}\partial_{\alpha}j^{\gamma}\cdot
\partial_{\beta}\left(  \partial_{\gamma}\theta+i\partial_{z}K\cdot
\partial_{\gamma}z-i\partial_{\overline{z}}K\cdot\partial_{\gamma}\overline
{z}\right)  . \label{nonc-e}%
\end{align}
The above relations show that the action (\ref{nonc-action}) is the
generalization of the perfect fluid to the noncommutative case because the
equations (\ref{em-tensor-fluid}), (\ref{nonc-p}) and (\ref{nonc-e}) reduce in
the limit $\lambda^{\alpha\beta}\rightarrow0$ to the known relations for the
energy-momentum tensor, the pressure and the energy density
\cite{Nyawelo:2003bv}. The pressure is the generalization of the Legendre
transformation of the specific energy to the noncommutative fluid. The
divergenceless of the energy-momentum tensor is aparent in the equation
(\ref{em-tensor-fluid}) from which we note the last term that involves the
product between the velocity and the combination of potentials that include
the nonzero vorticity. This resembles a dissipative term that is a consequence
of the noncommutative structure of the spacetime. If we require
that all momentum density be generated by the flow of the energy density, it
follows that%
\begin{equation}
\lambda^{\alpha\beta}j^{\mu}\partial_{\alpha}j_{\mu}\cdot\partial_{\beta
}\left(  \partial_{\nu}\theta+i\partial_{z}K\cdot\partial_{\nu}z-i\partial
_{\overline{z}}K\cdot\partial_{\nu}\overline{z}\right)  =0.
\label{mom-density}%
\end{equation}
If the fluid is generalized to include more conserving charges, one could use
the equation (\ref{mom-density}) to define $u^{\mu}$ which is the analogue of
choosing the frame for the commutative fluid.

\section{Volume preserving symmetry}

The noncommutative structure of spacetime given by equation (\ref{nc-rel}) is
invariant under the following generalization of the volume preserving
transformations \cite{lect-jackiw}%
\begin{equation}
\delta x_{\mu}=\left[  x_{\mu},h\right]  , \label{vol-pres-trans}%
\end{equation}
where the parameter $h(x)$ is an arbitrary continuos function on $x^{\mu}$'s.
The brackets from the above equation involve the Moyal product and at the
first order in $\lambda^{\mu\nu}$ take the form
\begin{equation}
\left[  f,g\right]  =i\lambda^{\mu\nu}\partial_{\mu}f\cdot\partial_{\nu}g.
\label{brackets}%
\end{equation}
In general, the Lagrangian given in the relation (\ref{nonc-lagrangian}) is
not invariant under the transformations (\ref{vol-pres-trans}) due to the
arbitrarieness of the functions $\theta(x)$, $z(x)$, $\bar{z}(x)$,
$K(z,\bar{z})$ and $f(x)$. Thus, by requiring that the Lagrangian be invariant
under the volume preserving transformations constraints need to be imposed on
these functions. It can be easily verified that the fields of the theory
transform under (\ref{vol-pres-trans}) as follows%
\begin{align}
\delta\phi &  =\left[  \phi,h\right]  ,\nonumber\\
\delta\psi^{\mu}  &  =\left[  \psi^{\mu},h\right]  ,\nonumber\\
\delta\left(  \partial_{\mu}\phi\right)   &  =\left[  \partial_{\mu}%
\phi,h\right]  +\left[  \phi,\partial_{\mu}h\right]  ,
\label{vol-pres-trans-field-1}%
\end{align}
where $\phi$ and $\psi^{\mu}$ are scalar and vector fields, respectively. The
transformation of the derivative holds for vector fields, too. By varying the
Lagrangian (\ref{nonc-lagrangian}) with respect to (\ref{vol-pres-trans}), one
obtains a bi-polynomial in the powers $m$\ of the antisymmetric matrix
$\lambda^{\mu\nu}$ and the degree $n$ of the derivatives of the arbitrary
parameter $h(x)$. Consequently, the invariance of the Lagrangian is guaranteed
if the terms of its variation vanish at each order in $m$ and $n$,
respectively. By keeping in mind this organization, we obtain from the terms
linear in $\lambda^{\mu\nu}$ the following equations%
\begin{align}
f^{\prime}\frac{j^{\mu}\partial_{\alpha}j_{\mu}}{\sqrt{-j^{2}-\frac{i}%
{2}\lambda^{\beta\gamma}\partial_{\beta}j^{\nu}\partial_{\gamma}j_{\nu}}}  &
=-\partial_{\alpha}\left[  j^{\mu}\left(  \partial_{\mu}\theta+i\partial
_{z}K\cdot\partial_{\mu}z-i\partial_{\overline{z}}K\cdot\partial_{\mu
}\overline{z}\right)  \right]  ,\label{const-lambda1-h1}\\
j^{\mu}\left(  \partial_{\alpha}\theta+i\partial_{z}K\cdot\partial_{\alpha
}z-i\partial_{\overline{z}}K\cdot\partial_{\alpha}\overline{z}\right)   &  =0.
\label{const-lambda1-h2}%
\end{align}
The quadratic terms in $\lambda^{\mu\nu}$ involve second and third order
partial derivatives of $h$. The second order derivatives couple with
$\lambda^{\mu\nu}$ as well as $j^{\mu}$ and different couplings generate
independent constraints. The result is the following set of equations%
\begin{align}
&  \partial_{\gamma}j^{\mu}\partial_{\alpha}\left(  \partial_{\mu}%
\theta+i\partial_{z}K\cdot\partial_{\mu}z-i\partial_{\overline{z}}%
K\cdot\partial_{\mu}\overline{z}\right)  +\partial_{\alpha}j^{\mu}%
\partial_{\gamma}\left(  \partial_{\mu}\theta+i\partial_{z}K\cdot\partial
_{\mu}z-i\partial_{\overline{z}}K\cdot\partial_{\mu}\overline{z}\right)
\nonumber\\
&  +2j^{\mu}\left[  \partial_{\lbrack\alpha}\partial_{z}K\cdot\partial
_{\gamma]}\partial_{\mu}z-\left(  z\leftrightarrow\bar{z}\right)  \right]
=0,\label{const-lambda2-h2-1}\\
&  \partial_{\alpha}j^{\mu}\left[  \partial_{\gamma}\partial_{\beta}%
\theta+i\partial_{\beta}\partial_{z}K\cdot\partial_{\gamma}z+i\partial
_{z}K\cdot\partial_{\gamma}\partial_{\beta}z-\left(  z\leftrightarrow\bar
{z}\right)  \right]  +j^{\mu}\left[  \partial_{\alpha}\partial_{z}%
K\cdot\partial_{\gamma}\partial_{\mu}z-\left(  z\leftrightarrow\bar{z}\right)
\right]  =0,\label{const-lambda2-h2-2}\\
&  \partial_{\alpha}j^{\mu}\left(  \partial_{\gamma}\theta+i\partial_{z}%
K\cdot\partial_{\gamma}z-i\partial_{\overline{z}}K\cdot\partial_{\gamma
}\overline{z}\right)  +j^{\mu}\left(  \partial_{\alpha}\partial_{z}%
K\cdot\partial_{\gamma}z-\partial_{\alpha}\partial_{\bar{z}}K\cdot
\partial_{\gamma}\bar{z}\right)  =0. \label{const-lambda2-h3}%
\end{align}
where we have used the standard antisymmetrization convention with respect to
the spacetime indices $a_{[\mu}b_{\nu]}=\frac{1}{2}\left(  a_{\mu}b_{\nu
}-a_{\nu}b_{\mu}\right)  $. Constraints with higher powers of $\lambda^{\mu
\nu}$ arise from higher order corrections to the Lagrangian. If the spacetime
noncommutativity is assumed to hold at high energy, only the linear terms in
the antisymmetric matrix are relevant to the theory and the invariance of the
Lagrangian under the generalized volume transformations is determined by the
constraints (\ref{const-lambda1-h1}) and (\ref{const-lambda1-h2}) alone. Also,
if the theory is studied on-shell, some simplification of the above set of
constraints is obtained.

\section{A simpler model}

The noncommutative perfect fluids discussed in the previous sections form a
general class since the functions $K(z,\bar{z})$ and $f(\sqrt{-j^{\mu}\ast
j_{\mu}})$ are not required to satisfy any property other than
differentiability to an arbitrary order. This makes the dynamics quite
complicate, even at first order in the noncommutative parameter. A slighty
simpler model can be obtained by taking%
\begin{equation}
K(z,\bar{z})=z\ast\bar{z},\quad f(\sqrt{-j^{\mu}\ast j_{\mu}})=\frac{c}{2}%
\rho^{2}=-\frac{c}{2}j^{2}, \label{K-f-example}%
\end{equation}
where $c$ is a c-number constant. In this model, the function $K(z,\bar{z})$
represents the generalization of the K\"{a}hler potential of the complex plane
and, at the first order in the noncommutative parameter, it is a noncomutative
deformation of the complex plane. The particular form of the function $f$ is
typical to the perfect fluid. The lagrangian (\ref{nonc-lagrangian}) of this
particular model can be casted in the following form at first order 
$\lambda^{\alpha \beta}$%
\begin{align}
\mathcal{L}=  &  -j^{\mu}\left(  \partial_{\mu}\theta+i\bar{z}\cdot
\partial_{\mu}z-iz\cdot\partial_{\mu}\bar{z}\right)  +\frac{1}{2}%
\lambda^{\alpha\beta}j^{\mu}\left(  \partial_{\alpha}\bar{z}\cdot
\partial_{\beta}\partial_{\mu}z-\partial_{\alpha}z\cdot\partial_{\beta
}\partial_{\mu}\bar{z}\right) \nonumber\\
&  -\frac{i}{2}\lambda^{\alpha\beta}\partial_{\alpha}j^{\mu}\cdot
\partial_{\beta}\left(  \partial_{\mu}\theta+i\bar{z}\cdot\partial_{\mu
}z-iz\cdot\partial_{\mu}\bar{z}\right)  +\frac{c}{2}j^{2}. \label{nonc-act-ex}%
\end{align}
The equations of motion can be obtained by using the relations
(\ref{K-f-example}) into the general equations (\ref{eq-mot-theta}%
)-(\ref{eq-mot-z-bar}) or by recalculating them from the scratch%
\begin{align}
\partial_{\mu}j^{\mu}=0  &  ,\label{em-ex-theta}\\
cj_{\mu}-\left(  \partial_{\mu}\theta+i\overline{z}\cdot\partial_{\mu
}z-iz\cdot\partial_{\mu}\overline{z}\right)  +\frac{1}{2}\lambda^{\alpha\beta
}\left(  \partial_{\alpha}\overline{z}\cdot\partial_{\beta}\partial_{\mu
}z-\partial_{\alpha}z\cdot\partial_{\beta}\partial_{\mu}\overline{z}\right)
=0  &  ,\label{em-ex-j}\\
j^{\mu}\partial_{\mu}\bar{z}=j^{\mu}\partial_{\mu}z=0  &  .
\label{em-ex-z-z-bar}%
\end{align}
The first remark that one can make about the dynamics of this particular model
is that the equations of motion of $\theta$, $z$ and $\bar{z}$ potentials do
not receive any noncommutative correction. Next, we note that the relations
(\ref{em-ex-z-z-bar}) imply the existence of an infinite set of currents%
\begin{equation}
J_{\mu}\left[  G\right]  =-2G(z,\bar{z})\cdot j_{\mu}, \label{curr-ex}%
\end{equation}
where the generators $G(z,\bar{z})$ are arbitrary commutative functions on
their arguments. The currents $J_{\mu}\left[  G\right]  $ are divergenceless
at zero order in the noncommutative parameter because at this order the
Leibniz rule holds. To the currents (\ref{curr-ex}) correspond the conserved
charges%
\begin{equation}
Q[G]=\int d^{3}xJ^{0}[G]. \label{ch-ex}%
\end{equation}
These properties show that the particular model described by the functions
(\ref{K-f-example}) shares similar properties with the whole class of the
commutative relativistic perfect fluids and with a special regime of the
supersymmetric fluids \cite{Nyawelo:2003bv,Holender:2008qj}.

Next, we can particularize the constraints (\ref{const-lambda1-h1}) -
(\ref{const-lambda2-h3}) on the field potentials under which the Lagrangian
(\ref{nonc-act-ex}) becomes invariant under the volume preserving symmetry. If
we consider the on-shell invariance, then the constraints take the simpler
form%
\begin{align}
&  cj^{\mu}\partial_{\alpha}j_{\mu}+\partial_{\alpha}\left(  j^{\mu}%
\partial_{\mu}\theta\right)  =0,\label{c-ex-1}\\
&  j^{\mu}\left(  \partial_{\alpha}\theta+i\bar{z}\partial_{\alpha
}z-iz\partial_{\alpha}\overline{z}\right)  =0,\label{c-ex-2}\\
&  \partial_{\gamma}j^{\mu}\partial_{\alpha}\left(  \partial_{\mu}\theta
+i\bar{z}\partial_{\mu}z-iz\partial_{\mu}\overline{z}\right)  +\partial
_{\alpha}j^{\mu}\partial_{\gamma}\left(  \partial_{\mu}\theta+i\bar{z}%
\partial_{\mu}z-iz\partial_{\mu}\overline{z}\right) \nonumber\\
&  +2j^{\mu}\left[  \partial_{\lbrack\alpha}\bar{z}\partial_{\gamma]}%
\partial_{\mu}z-\left(  z\leftrightarrow\bar{z}\right)  \right]
=0,\label{c-ex-3}\\
&  \partial_{\alpha}j^{\mu}\left[  \partial_{\gamma}\partial_{\beta}%
\theta+i\partial_{\beta}\bar{z}\cdot\partial_{\gamma}z+i\bar{z}\cdot
\partial_{\gamma}\partial_{\beta}z-\left(  z\leftrightarrow\bar{z}\right)
\right]  +j^{\mu}\left[  \partial_{\alpha}\bar{z}\cdot\partial_{\gamma
}\partial_{\mu}z-\left(  z\leftrightarrow\bar{z}\right)  \right]
=0,\label{c-ex-4}\\
&  \partial_{\alpha}j^{\mu}\left(  \partial_{\gamma}\theta+i\bar{z}%
\partial_{\gamma}z-iz\partial_{\gamma}\overline{z}\right)  +j^{\mu}\left(
\partial_{\alpha}\bar{z}\cdot\partial_{\gamma}z-\partial_{\alpha}%
z\cdot\partial_{\gamma}\bar{z}\right)  =0. \label{c-ex-5}%
\end{align}
The fluid properties of the model are described by the energy-momentum tensor
and the equation of state which can be easily obtained from the equations
(\ref{em-tensor-fluid}) - (\ref{nonc-e}) and put into the following form%
\begin{equation}
T_{\mu\nu}=\eta_{\mu\nu}p(\lambda)+\left[  \varepsilon(\lambda)+p(\lambda
)\right]  u_{\mu}u_{\nu}+i\lambda^{\alpha\beta}\partial_{\alpha}\rho\cdot
u_{\mu}\partial_{\beta}\left(  \partial_{\nu}\theta+i\overline{z}\partial
_{\nu}z-iz\partial_{\nu}\overline{z}\right)  , \label{emt-ex}%
\end{equation}
where%
\begin{align}
p(\lambda)  &  =\rho f^{\prime}-f-j^{\mu}\partial_{\mu}\theta+\frac{1}%
{2}\lambda^{\alpha\beta}j^{\mu}\left(  \partial_{\alpha}\overline{z}%
\cdot\partial_{\beta}\partial_{\mu}z-\partial_{\alpha}z\cdot\partial_{\beta
}\partial_{\mu}\bar{z}\right) \nonumber\\
&  -\frac{i}{2}\lambda^{\alpha\beta}\partial_{\alpha}j^{\mu}\cdot
\partial_{\beta}\left(  \partial_{\mu}\theta+i\overline{z}\cdot\partial_{\mu
}z-iz\cdot\partial_{\mu}\overline{z}\right)  ,\label{pp-ex}\\
\varepsilon(\lambda)  &  =f+j^{\mu}\partial_{\mu}\theta-\frac{1}{2}%
\lambda^{\alpha\beta}j^{\mu}\left(  \partial_{\alpha}\overline{z}\cdot
\partial_{\beta}\partial_{\mu}z-\partial_{\alpha}z\cdot\partial_{\beta
}\partial_{\mu}\bar{z}\right) \nonumber\\
&  +\frac{i}{2}\lambda^{\alpha\beta}\partial_{\alpha}j^{\mu}\cdot
\partial_{\beta}\left(  \partial_{\mu}\theta+i\overline{z}\partial_{\mu
}z-iz\partial_{\mu}\overline{z}\right)  . \label{e-ex}%
\end{align}
From these equations, we see that the present model represents a
generalization of the relativistic perfect fluid which preserves the infinite
conserved currents associated with the reparametrization invariance of the
complex manifold which is described by the complex potentials $z$ and $\bar{z}$ 
at zeroth order in the noncommutative parameter.

\section{Conclusions and Discussions}

In this paper, we have proposed the functional action (\ref{nonc-action}) for
a large class of noncommutative fluids that generalizes the relativistic
perfect fluids formulated in the K\"{a}hler parametrization to the
noncommutative spacetime. The noncommutative fluids are characterized by 
$K(z,z)$ and $f(\sqrt{-j^{\mu}\ast j_{\mu}})$ which generalize the 
corresponding arbitrary functions from the commutative case with
the restriction of the partial derivatives to the zeroth order in the
noncommutative parameter that makes the Leibniz property hold. Without this
technical restriction, there are more contributions at first order in
$\lambda^{\mu\nu}$. We have derived the equations of motion of the fluid
potentials to the first order in the noncommutative parameter. Also, we have
calculated the energy-momentum tensor. The equation of motion for the $\theta$
- field (\ref{eq-mot-theta}) does not receive any noncommutative corrections
and it represents the divergenceless of the density current $j^{\mu}$\ like in
the commutative case. However, the energy-momentum tensor is not
divergenceless. That implies that $T_{\mu\nu}$ is not invariant under
translations if the dual operators $P_{\mu}=\partial_{\mu}$ commute with each
other. If one requires that the energy-momentum tensor of the noncommutative
theory be invariant, the constraints (\ref{cons-em-tensor-1}) and
(\ref{cons-em-tensor-2}) should be imposed on the fields. Note that the
equation (\ref{eq-mot-theta}) holds in the $\kappa$ - Minkowski spacetime,
too. Actually, the current conservation suggests that the action
(\ref{nonc-action}) be valid in all noncommutative spaces where the
translations are generated by commuting $P_{\mu}=\partial_{\mu}$. The equation
of motion of the current $j^{\mu}$ contains commutative terms that are the
same as the ones obtained for commutative fluids and noncommutative
corrections. Also, one can show that the equations (\ref{eq-mot-z}) and
(\ref{eq-mot-z-bar}) for the fields $z$ and $\bar{z}$\ can be reduced to the
corresponding equations in the commutative case if the current conservation
(\ref{eq-mot-theta}) is used in those terms that are independent of
$\lambda^{\mu\nu}$. By particulazing the functions $K(z,z)$ and $f$ to the
relations (\ref{K-f-example}), we have shown that other properties of the the
commutative fluids can be generalized to the noncommutative ones. In
particular, the models especified by (\ref{K-f-example}) have an infinity of
conserved currents $J_{\mu}\left[  G\right]  $ in the Leibniz approximation
for the partial derivatives. This feature alone makes the model quite
interesting, since in general the currents are not conserved for
generalizations of the perfect fluid.

Another important quantity that is conserved in the commutative case is the
axial current which is related to the topologically conserved linking number
of vortices \cite{Nyawelo:2003bv}. Therefore, it is desirable to see if the
noncommutative fluids have divergenceless axial currents. We can generalize
the axial current $k_{\mu}$\ to the noncommutative case by applying the
correspondence principle adopted in this paper%
\begin{equation}
K^{\mu}=\epsilon^{\mu\nu\xi\lambda}\left(  \partial_{\nu}\theta+i\partial
_{z}K\ast\partial_{\nu}z-i\partial_{\overline{z}}K\ast\partial_{\nu}%
\overline{z}\right)  \ast\partial_{\xi}\left(  \partial_{\lambda}%
\theta+i\partial_{z}K\ast\partial_{\lambda}z-i\partial_{\overline{z}}%
K\ast\partial_{\lambda}\overline{z}\right)  , \label{nc-axial-cur}%
\end{equation}
where $\epsilon^{\mu\nu\xi\lambda}$ is the four-dimensional antisymmetric
tensor with $\epsilon^{0123}=1$. If we calculate the divergence of $K^{\mu}$
at first order in $\lambda^{\mu\nu}$ we see, after lenghty calculations, that
it fails to be zero by a term of the form%
\begin{align}
&  -2i\epsilon^{\mu\nu\xi\lambda}\lambda^{\alpha\beta}\left(  \partial
_{z\overline{z}}^{2}K\partial_{\mu}\overline{z}\partial_{\alpha}\partial_{\nu
}z+\partial_{\overline{z}}\partial_{z}^{2}K\partial_{\mu}\overline{z}%
\partial_{\alpha}z\partial_{\nu}z+\partial_{\overline{z}}^{2}\partial
_{z}K\partial_{\mu}\overline{z}\partial_{\alpha}\overline{z}\partial_{\nu
}z+\partial_{z\overline{z}}^{2}K\partial_{\mu}\partial_{\alpha}\overline
{z}\partial_{\nu}z\right) \nonumber\\
&  \times\left(  \partial_{z\overline{z}}^{2}K\partial_{\xi}\overline
{z}\partial_{\beta}\partial_{\lambda}z+\partial_{\overline{z}}\partial_{z}%
^{2}K\partial_{\xi}\overline{z}\partial_{\beta}z\partial_{\lambda}%
z+\partial_{\overline{z}}^{2}\partial_{z}K\partial_{\xi}\overline{z}%
\partial_{\beta}\overline{z}\partial_{\lambda}z+\partial_{z\overline{z}}%
^{2}K\partial_{\xi}\partial_{\beta}\overline{z}\partial_{\lambda}z\right)
\nonumber\\
&  -i\epsilon^{\mu\nu\xi\lambda}\lambda^{\alpha\beta}\partial_{z\overline{z}%
}K\partial_{\mu}\overline{z}\partial_{\nu}z\left(  \partial_{z}^{2}%
K\partial_{\xi}\partial_{\alpha}z\partial_{\beta}\partial_{\lambda}%
z-\partial_{\overline{z}}^{2}K\partial_{\xi}\partial_{\alpha}\overline
{z}\partial_{\beta}\partial_{\lambda}\overline{z}\right)  .
\label{nc-axial-div}%
\end{align}
This relation shows that $K^{\mu}$ would not be conserved unless further
constraints were imposed on the potentials. However, we can show that for the
particular model presented in the section 5%
\begin{equation}
\partial_{\mu}K^{\mu}=0. \label{nc-axial-cons}%
\end{equation}
Thus, the generalization of the K\"{a}hler potential for the complex plane and
the perfect fluid shares most of the properties with the commutative fluids.

It is interesting to investigate further the noncommutative fluids of the type
presented in this paper along several lines. One of the most important
problems is to describe concrete models that preserve the noncommutative
Poincar\'{e} symmetry. This can be achieved by taking for $M_{\lambda}$ the
$\kappa$ - Minkowski spacetime. As mentioned above, the noncommutative
generalization of the translation operators satisfy the equation
(\ref{nc-rel}) so all the conclussions derived for it concerning the
invariance of the density current and the energy-momentum tensor are expected
to continue true. Another interesting issue is to analyse the fluids obtained
by relaxing the Leibniz rule for the partial derivative and work within the
full noncommutative structure. This would modify all the equations of motion
and the constraints by adding extra terms that contain $\lambda^{\mu\nu}$.
Therefore, one should be able to recover the relativistic fluid in the
commutative limit as we have done in the present paper. However, the
conservation of the generalized parametrization currents might not hold
without other constraints. And finally, it would be interesting to study the
symplectic structure on the phase space of the fluid induced by the underlying
noncommutative structure of spacetime.

\noindent\textbf{Acknowledgements} I. V. V. would like to M. D. Roberts for
correspondence and to S. V. de Borba Gon\c{c}alves for hospitality at
PPGFis-UFES where part of this work was accomplished. L. H. acknowledges the
support of CAPES/Prodoctoral programme.

%\newpage
%\begin{appendix}
%\section{Nonequilibrium Thermo Field Dynamics}
%\end{appendix}

%\newpage

\end{document}